\documentclass[aps,twocolumn]{revtex4-1}

\usepackage{graphicx}
\usepackage{natbib}
\usepackage[dvipsnames]{xcolor}

\usepackage{graphics}
\usepackage{epsfig}
\usepackage{bm}
\usepackage{amsmath,amssymb}
\usepackage{stmaryrd}
\usepackage{wasysym}

\def\vb#1{\mbox{\boldmath$#1$}}
\def\pd#1#2{\frac{\partial #1}{\partial #2}}
\def\fd#1#2{\frac{\delta #1}{\delta #2}}
\def\wh#1{\widehat{#1}}
\def\bdot{\,\vb{\cdot}\,}
\def\btimes{\,\vb{\times}\,}

\def\bhat{\wh{{\sf b}}}
\def\cal#1{\mathcal{#1}}

\def\exd{{\sf d}}
\def\bhat{\wh{{\sf b}}}

\newcommand{\bc}{\begin{center}}
\newcommand{\ec}{\end{center}}
\newcommand{\bt}{\begin{tabbing}}
\newcommand{\et}{\end{tabbing}}
\newcommand{\be}{\begin{equation}}
\newcommand{\ee}{\end{equation}}
\newcommand{\ba}{\begin{eqnarray}}
\newcommand{\ea}{\end{eqnarray}}

\begin{document}

\title{Hamiltonian Structure of the Guiding-center Vlasov-Maxwell Equations \\ with Polarization and Magnetization}

\author{Alain J.~Brizard$^{a}$}
\affiliation{Department of Physics, Saint Michael's College, Colchester, VT 05439, USA \\ $^{a}$Author to whom correspondence should be addressed: abrizard@smcvt.edu}

\begin{abstract}
The Hamiltonian formulation of guiding-center Vlasov-Maxwell equations, which contain dipole contributions to the guiding-center polarization and magnetization, is presented in terms of a guiding-center Hamiltonian functional that is derived from the exact guiding-center Vlasov-Maxwell energy conservation law, and an antisymmetric functional bracket that satisfies the Jacobi property. Exact energy-momentum and angular momentum conservation laws are expressed in Hamiltonian form and the guiding-center Vlasov-Maxwell entropy functional is shown to be a Casimir functional.
\end{abstract}

\date{\today}

\maketitle

\section{Introduction}

The investigation of the noncanonical Hamiltonian structure of important dissipation's plasma dynamical systems began with the equations of ideal magnetohydrodynamics (MHD) \cite{Morrison_Greene_1980} and the Vlasov-Maxwell equations \citep{Morrison_1980,Morrison_1982,Marsden_Weinstein_1982}. These structures have led to the development of advanced structure-preserving algorithms \citep{Evstatiev_Shadwick_2013,Kraus_2017,Morrison_2017,Xiao_Qin_2018,Kormann_2021} that allow for sophisticated numerical methods for solving dissipationless plasma equations that possess Hamiltonian and Lagrangian structures.

Based on the asymptotic separation of orbital time scales in a magnetically-confined plasma \cite{Northrop_1963,RGL_1981,RGL_1983,Cary_Brizard_2009}, reduced Vlasov-Maxwell equations have played a crucial role in the analysis of the long-time behavior of magnetized plasmas \cite{Brizard_Hahm_2007,Brizard_2008,Cary_Brizard_2009}. On the one hand, through the introduction of adiabatic invariants, constructed by standard Lie-transform perturbation methods \citep{RGL_1982} associated with widely-separated orbital time scales, fast orbital time scales are asymptotically removed from the Vlasov kinetic equation to produce a reduced Vlasov equation. On the other hand, this dynamical reduction introduces important polarization and magnetization effects into the reduced Maxwell equations \citep{Brizard_2008}. In addition, the variational structure of these reduced Vlasov-Maxwell equations allows the derivation of exact energy-momentum conservation laws through the application of the Noether method \citep{Brizard_2008}.

The purpose of the present paper is to explore the Hamiltonian structure of the guiding-center Vlasov-Maxwell equations. One approach is to proceed by Lie-transform methods \citep{Brizard_Lie_2016} on the Hamiltonian structure of the Vlasov-Maxwell equations \cite{Morrison_1980,Morrison_1982,Marsden_Weinstein_1982}. A more direct approach, however, is to construct the Hamiltonian structure from the guiding-center Vlasov-Maxwell equations themselves. (The construction of the Hamiltonian structure for the Vlasov-Maxwell equations followed by Morrison \cite{Morrison_1982} is reviewed in App.~\ref{sec:VM}.) This direct approach yielded the Hamiltonian structure for the guiding-center Vlasov-Maxwell equations \cite{Brizard_2021_gcVM} in the simplest case when the guiding-center polarization appears only as a correction to the guiding-center magnetization \citep{Brizard_Tronci_2016}. A direct approach also yielded the Hamiltonian structure for the gauge-free gyrokinetic Vlasov-Maxwell theory \cite{Brizard_2021_gyVM}, which was made possible only because the gyrokinetic equations were expressed solely in terms of the perturbed electromagnetic fields $({\bf E}_{1},{\bf B}_{1})$, instead of the standard gyrokinetic representation \cite{Brizard_Hahm_2007} involving the perturbed electromagnetic potentials $(\Phi_{1},{\bf A}_{1})$.

In order to take into account the full guiding-center polarization, it is a standard procedure \cite{Pfirsch_1984,Pfirsch_Morrison_1985,Brizard_1995,Brizard_2024} to describe particle motion in a local frame moving with the $E\times B$ velocity, expressed as ${\bf E}\btimes c\bhat/B$ (in terms of the electric field ${\bf E}$ and the magnetic field ${\bf B} \equiv B\,\bhat$). In this moving frame, a generic expression for the guiding-center Lagrangian is expressed as \cite{Pfirsch_1984,Pfirsch_Morrison_1985,Brizard_1995,Brizard_2024}
\begin{equation}
L_{\rm gc} \;\equiv\; \left(\frac{e}{c}\,{\bf A} + P_{\|}\,\bhat + \vb{\Pi}_{\rm gc}\right)\bdot\dot{\bf X} \;-\; \left(e\,\Phi \;+\frac{}{} K_{\rm gc}\right),
\label{eq:Lgc_def}
\end{equation}
where $({\bf X},P_{\|})$ denote the guiding-center position and guiding-center parallel kinetic momentum, while the guiding-center {\it symplectic} momentum $\vb{\Pi}_{\rm gc}$ and the guiding-center kinetic energy $K_{\rm gc}$ may depend on the electric field ${\bf E}$. Hence, in this {\it symplectic} polarization representation, each term separately contributes to guiding-center polarization, which is defined in terms of the partial derivative of Eq.~\eqref{eq:Lgc_def} with respect to ${\bf E}$:
\begin{equation}
\partial L_{\rm gc}/\partial{\bf E} \;\equiv\; \dot{\bf X} \bdot\partial\vb{\Pi}_{\rm gc}/\partial{\bf E} \;-\; \partial K_{\rm gc}/\partial{\bf E}. 
\label{eq:Lag_E}
\end{equation}
In addition, we note that, despite the fact that the electromagnetic potentials $(\Phi,{\bf A})$ appear in the guiding-center Lagrangian \eqref{eq:Lgc_def}, the guiding-center equations of motion derived from it are gauge invariant since Eq.~\eqref{eq:Lgc_def} transforms as $L_{\rm gc} \rightarrow L_{\rm gc} + (e/c)\,d\chi/dt$ under the gauge transformation $(\Phi,{\bf A}) \rightarrow (\Phi - c^{-1}\partial\chi/\partial t,\; {\bf A} + \nabla\chi)$, and the guiding-center Euler-Lagrange equations are invariant under the addition of an exact time derivative to the Lagrangian $L_{\rm gc}$.

In the present paper, we adopt the alternate {\it Hamiltonian} polarization representation in which the guiding-center polarization is introduced explicitly only through the guiding-center kinetic energy $K_{\rm gc}$, i.e., Eq.~\eqref{eq:Lag_E} simplifies to $\partial L_{\rm gc}/\partial{\bf E} \equiv -\,\partial K_{\rm gc}/\partial{\bf E}$ when $\partial\vb{\Pi}_{\rm gc}/\partial{\bf E} \equiv 0$. These dual polarization representations are analogous to the Hamiltonian and symplectic representations of gyrokinetic Vlasov-Maxwell theory \cite{HLB_1988,Brizard_1989,Brizard_Hahm_2007}, in which the perturbed magnetic field, respectively, appears either in the gyrocenter Hamiltonian only or in both the gyrocenter Hamiltonian and the gyrocenter Poisson bracket \cite{HLB_1988,Brizard_1989}. We note that, in the gyrocenter and guiding-center symplectic representations, the inductive electric field and the polarization drift velocity appear explicitly in their respective Euler-Lagrange equations, while these terms are absent in the Hamiltonian representations. Hence, because of the simplicity of the Hamiltonian polarization representation, we will explore the Hamiltonian structure of the guiding-center Vlasov-Maxwell equations derived from the guiding-center Lagrangian \eqref{eq:Lgc_def} with $\vb{\Pi}_{\rm gc} \equiv 0$ used for further simplicity, and leave the symplectic polarization case to future work.

The remainder of the paper is organized as follows. In Sec.~\ref{sec:gc_var}, the guiding-center Vlasov-Maxwell equations are presented, with the guiding-center polarization and magnetization defined in terms of derivatives of the guiding-center Lagrangian with respect to electric and magnetic fields, respectively. In Sec.~\ref{sec:gc_Noether}, the energy-momentum conservation laws of the guiding-center Vlasov-Maxwell equations, which are derived by Noether method, are presented. In addition, we show that the angular momentum conservation law follows from the explicit symmetry of the guiding-center stress tensor. In Sec.~\ref{sec:Ham_gcVM}, we construct the functional bracket for the guiding-center Vlasov-Maxwell equations, following a direct procedure adopted for other reduced Vlasov-Maxwell equations \cite{Brizard_2021_gcVM,Brizard_2021_gyVM}. The form of the new guiding-center functional bracket will guarantee the Jacobi property.
 
\section{\label{sec:gc_var}Guiding-center Vlasov-Maxwell equations}

The guiding-center Vlasov-Maxwell equations describe the time evolution of the guiding-center Vlasov phase-space density $F_{\rm gc} = {\cal J}_{\rm gc}\,F$, which is defined in terms of the guiding-center Vlasov function $F({\bf X},P_{\|},J,t)$ and the guiding-center phase-space Jacobian ${\cal J}_{\rm gc}$ on a $4+1$ reduced guiding-center phase space $({\bf X},P_{\|};J)$, where $J \equiv \mu B/\Omega = \mu\,(mc/e)$ denotes the guiding-center gyroaction adiabatic invariant, and the Maxwell fields $({\bf D}_{\rm gc}, {\bf B})$, where the guiding-center displacement field ${\bf D}_{\rm gc} \equiv {\bf E} + 4\pi\,\vb{\sf P}_{\rm gc}$ is defined in terms of the guiding-center polarization $\vb{\sf P}_{\rm gc}$, while ${\bf B}$ denotes the magnetic field. We note here that, in contrast to gyrokinetic Vlasov-Maxwell theory \cite{Brizard_Hahm_2007}, the electric and magnetic fields are not separated into background (unperturbed) and fluctuating (perturbed) fields, where only the fluctuating fields are considered as variational fields.

As discussed by Morrison \cite{Morrison_2013}, the functional bracket for the guiding-center Vlasov-Maxwell equations considered here will act on arbitrary functionals ${\cal F}[F_{\rm gc},{\bf D}_{\rm gc},{\bf B}]$ and, in the Hamiltonian representation adopted here, the electric field ${\bf E}[F_{\rm gc},{\bf D}_{\rm gc},{\bf B}]$ will be shown to be expressed as a simple functional of the guiding-center fields $(F_{\rm gc},{\bf D}_{\rm gc},{\bf B})$. Lastly, we will show that the guiding-center Vlasov-Maxwell equations 
\begin{eqnarray}
 \pd{F_{\rm gc}}{t} &=& -\;\nabla\bdot\left(F_{\rm gc}\frac{}{}\dot{\bf X}\right) \;-\; \pd{}{P_{\|}}\left(F_{\rm gc}\frac{}{}\dot{P}_{\|}\right),  \label{eq:gcVlasov_div} \\
 \pd{{\bf D}_{\rm gc}}{t} &=& c\,\nabla\btimes{\bf H}_{\rm gc} \;-\; 4\pi\;{\bf J}_{\rm gc}, \label{eq:gc_Maxwell} \\
 \pd{\bf B}{t} &=& -\,c\,\nabla\btimes{\bf E}, \label{eq:Faraday}
 \end{eqnarray}
can be expressed in Hamiltonian form
\begin{equation}
\pd{\cal F}{t} \;\equiv\; \left[{\cal F},\frac{}{} {\cal H}_{\rm gc}\right]_{\rm gc},
\label{eq:F_HB}
\end{equation}
where ${\cal H}_{\rm gc}$ denotes the guiding-center Vlasov-Maxwell Hamiltonian functional and $[\;,\;]_{\rm gc}$ denotes the guiding-center Hamiltonian functional bracket. In Eq.~\eqref{eq:gcVlasov_div}, the guiding-center equations of motion $(\dot{\bf X},
\dot{P}_{\|})$ will be derived (see App.~\ref{sec:App} for details) from the guiding-center Lagrangian 
\begin{eqnarray}
L_{\rm gc} &=&  \left(\frac{e}{c}\,{\bf A} \;+\; P_{\|}\;\bhat\right)\bdot\dot{\bf X} \;-\; \left( e\,\Phi \;+\frac{}{} K_{\rm gc} \right).
\label{eq:Lag_gc}
\end{eqnarray} 
In Eq.~\eqref{eq:gc_Maxwell}, on the other hand, the guiding-center electromagnetic fields
 \begin{eqnarray}
 {\bf D}_{\rm gc} &\equiv& {\bf E} \;+\; 4\pi\,\vb{\sf P}_{\rm gc}, \label{eq:D_gc} \\
 {\bf H}_{\rm gc} &\equiv& {\bf B} \;-\; 4\pi\,\vb{\sf M}_{\rm gc}, \label{eq:H_gc} 
\end{eqnarray}
are defined in terms of the guiding-center polarization and magnetization
\begin{equation}
\left(\vb{\sf P}_{\rm gc},\frac{}{} \vb{\sf M}_{\rm gc}\right) \;\equiv\; \int_{P} F_{\rm gc} \left(-\;\partial K_{\rm gc}/\partial{\bf E},\frac{}{} \partial L_{\rm gc}/\partial{\bf B}\right),
\label{eq:Pol_Mag_gc}
\end{equation}
where the guiding-center polarization is defined in the Hamiltonian representation, and we use the notation $\int_{P} (\cdots) = \sum\,\int (\cdots)\,2\pi dP_{\|} dJ$, which includes a summation over charged-particle species and excludes the guiding-center Jacobian 
${\cal J}_{\rm gc}$ since it is included in the definition of the guiding-center Vlasov density $F_{\rm gc}$. The guiding-center Vlasov-Maxwell equations \eqref{eq:gcVlasov_div}-\eqref{eq:Faraday} are complemented by the remaining Maxwell equations
\begin{eqnarray}
\nabla\bdot{\bf D}_{\rm gc} &=& 4\pi\,\varrho_{\rm gc}, \label{eq:Dgc_div} \\
\nabla\bdot{\bf B} &=& 0, \label{eq:B_div}
\end{eqnarray}
where the guiding-center charge and current densities
  \begin{equation}
  (\varrho_{\rm gc}, {\bf J}_{\rm gc}) \;=\; \int_{P} F_{\rm gc} \left(e,\frac{}{} e\,\dot{\bf X}\right),
  \label{eq:rho_J_gc}
  \end{equation}
 satisfy the charge conservation law 
\begin{equation}
\pd{\varrho_{\rm gc}}{t} \;+\; \nabla\bdot{\bf J}_{\rm gc} = 0,
\label{eq:charge_cons}
\end{equation}
which can be obtained directly from Eqs.~\eqref{eq:gc_Maxwell} and \eqref{eq:Dgc_div}.
 
\subsection{Guiding-center Hamilton equations}

According to the Hamiltonian polarization representation adopted here, the guiding-center kinetic energy  appearing in the guiding-center Lagrangian \eqref{eq:Lag_gc} is expressed as \citep{Cary_Brizard_2009} 
 \begin{equation}
 K_{\rm gc} \;=\; \mu\,B \;+\; \frac{P_{\|}^{2}}{2m} \;-\; \frac{m}{2}\,|{\bf E}\btimes c\bhat/B|^{2},
 \label{eq:K_gc_def}
  \end{equation}
which includes a ponderomotive contribution associated with the $E\times B$ velocity. The details of the guiding-center transformation leading to Eq.~\eqref{eq:K_gc_def} are presented in App.~\ref{sec:App}.

The Euler-Lagrange equations derived from the guiding-center Lagrangian \eqref{eq:Lag_gc} yield the guiding-center Hamilton equations of motion in reduced guiding-center phase space $({\bf X},P_{\|})$, which include the guiding-center velocity
 \begin{eqnarray}
 \dot{\bf X} &=& \pd{K_{\rm gc}}{P_{\|}}\;\frac{{\bf B}^{*}}{B_{\|}^{*}} \;+\; \left(e\,{\bf E}^{*} \;-\frac{}{} \nabla K_{\rm gc}\right)\btimes\frac{c\bhat}{eB_{\|}^{*}} \nonumber \\
  &\equiv& \{ {\bf X},\; K_{\rm gc}\}_{\rm gc} \;+\; e\,{\bf E}^{*}\bdot\{{\bf X},\; {\bf X}\}_{\rm gc},
 \label{eq:Xgc_dot}
 \end{eqnarray}
and the guiding-center parallel force equation
 \begin{eqnarray}
 \dot{P}_{\|} &=& \left(e\,{\bf E}^{*} \;-\frac{}{} \nabla K_{\rm gc}\right)\bdot\frac{{\bf B}^{*}}{B_{\|}^{*}}  \nonumber \\
  &\equiv&  \{ P_{\|},\; K_{\rm gc}\}_{\rm gc} \;+\; e\,{\bf E}^{*}\bdot\{{\bf X},\; P_{\|}\}_{\rm gc},
 \label{eq:Pgc_dot}
 \end{eqnarray}
 where $\bhat\bdot\dot{\bf X} = P_{\|}/m$ defines the parallel guiding-center velocity and the modified electric and magnetic fields are defined as
 \begin{eqnarray}
{\bf E}^{*} &\equiv& {\bf E} - \frac{P_{\|}}{e}\;\pd{\bhat}{t} = {\bf E} - \frac{P_{\|}}{eB}\,\left(\mathbb{I} - \bhat\bhat\right)\bdot\pd{\bf B}{t}, \label{eq:E_star} \\
{\bf B}^{*} &\equiv& {\bf B} \;+\; \frac{cP_{\|}}{e}\;\nabla\btimes\bhat. \label{eq:B_star}
  \end{eqnarray}
In accordance with standard guiding-center theory \cite{Kulsrud_1983}, we use the standard guiding-center ordering ${\bf E} = {\bf E}_{\bot} + \epsilon\,E_{\|}\,\bhat$, which holds that the parallel electric field is small compared with its components perpendicular to the magnetic field.

The reduced guiding-center Poisson bracket introduced in Eqs.~\eqref{eq:Xgc_dot}-\eqref{eq:Pgc_dot} is defined as
\begin{eqnarray}
\{F, \; G\}_{\rm gc} &=& \frac{{\bf B}^{*}}{B_{\|}^{*}}\bdot\left(\nabla F\,\pd{G}{P_{\|}} - \pd{F}{P_{\|}}\,\nabla G\right) \nonumber \\
 &&-\; \frac{c\bhat}{eB_{\|}^{*}}\bdot\nabla F\btimes\nabla G, 
   \label{eq:gcPB}
 \end{eqnarray}
 where $F$ and $G$ are arbitrary functions of the reduced guiding-center phase-space coordinates $Z^{\alpha} = ({\bf X},P_{\|})$. The guiding-center Jacobian ${\cal J}_{\rm gc} = (e/c)\,B_{\|}^{*}$, on the other hand, where $B_{\|}^{*} \equiv \bhat\bdot
 {\bf B}^{*}$, satisfies the guiding-center Liouville equation
\begin{equation}
\pd{{\cal J}_{\rm gc}}{t} \;+\; \nabla\bdot\left({\cal J}_{\rm gc}\frac{}{}\dot{\bf X}\right) \;+\; \pd{}{P_{\|}}\left({\cal J}_{\rm gc}\frac{}{}\dot{P}_{\|} \right) \;=\; 0,
\label{eq:gc_Liouville}
\end{equation}
where $(\dot{\bf X},\dot{P}_{\|})$ are given by Eqs.~\eqref{eq:Xgc_dot}-\eqref{eq:Pgc_dot}. As noted in App.~\ref{sec:VM}, the extended guiding-center Poisson bracket \eqref{eq:egc_PB} is not needed in what follows in order to guarantee the antisymmetry of the Hamiltonian functional bracket \eqref{eq:gcVM_bracket}.
 
 We note that the reduced guiding-center Poisson bracket \eqref{eq:gcPB} can be expressed in divergence form as
\begin{equation}
\{F, G\}_{\rm gc} \;=\; \frac{1}{B_{\|}^{*}}\pd{}{Z^{\alpha}}\left(B_{\|}^{*}\frac{}{}F\;\left\{ Z^{\alpha},\; G\right\}_{\rm gc}\right),
\label{eq:gcPB_div}
\end{equation}
and that it automatically satisfies the Jacobi identity expressed in terms of the Jacobiator
 \begin{eqnarray}
{\sf Jac}[F,G,K] &\equiv& \left\{F,\frac{}{}\{G, K\}_{\rm gc}\right\}_{\rm gc} \;+\;  \left\{G,\frac{}{}\{K,F\}_{\rm gc}\right\}_{\rm gc} \nonumber \\
 &&+\;  \left\{K,\frac{}{}\{F,G\}_{\rm gc}\right\}_{\rm gc} \;=\; 0,
 \label{eq:Jac_gcPB}
 \end{eqnarray}
which holds for any three arbitrary functions $(F,G,K)$ and follows from the condition $\nabla\bdot{\bf B}^{*} = \nabla\bdot{\bf B} \equiv 0$. 

  \subsection{\label{sec:Vlasov_gc}Guiding-center polarization and magnetization}
  
According to Eq.~\eqref{eq:Pol_Mag_gc}, the guiding-center polarization $\vb{\sf P}_{\rm gc}$ and magnetization $\vb{\sf M}_{\rm gc}$ can be derived from the guiding-center Lagrangian \eqref{eq:Lag_gc} by considering the electric and magnetic variations 
\begin{eqnarray}
\delta_{\bf E}L_{\rm gc} &=& -\; \delta_{\bf E}K_{\rm gc}, \label{eq:delta_E_Lag} \\
\delta_{\bf B}L_{\rm gc} &=& \delta_{\bf B}(P_{\|}\;\bhat)\bdot\dot{\bf X} - \delta_{\bf B}K_{\rm gc}, \label{eq:delta_B_Lag}
\end{eqnarray}
where the guiding-center velocity $\dot{\bf X}$ is kept constant. Here, the variation of the guiding-center symplectic momentum
\[ \delta_{\bf B}(P_{\|}\;\bhat) \;=\; (P_{\|}/B)\;\left(\mathbb{I} - \bhat\bhat\right)\bdot\delta{\bf B} \]
yields the moving electric-dipole contribution to the guiding-center magnetization 
\begin{eqnarray}
\delta_{\bf B}(P_{\|}\;\bhat)\bdot\dot{\bf X} &=& \dot{\bf X}\bdot \frac{P_{\|}}{B}\left(\mathbb{I} - \bhat\bhat\right)\bdot\delta{\bf B} = \left(\frac{e}{c}\dot{\bf X}\bdot\mathbb{P}_{\|}\right)\bdot\delta{\bf B} \nonumber \\
  &\equiv&  \left(\vb{\pi}_{\rm gc}\btimes P_{\|}\bhat/mc\right)\bdot\delta{\bf B},
  \label{eq:deltaB_P}
\end{eqnarray}
where 
\begin{equation}
\vb{\pi}_{\rm gc} \equiv (e\bhat/\Omega)\btimes\dot{\bf X}
\label{eq:Pi_gc}
\end{equation}
denotes the guiding-center electric dipole moment \citep{Brizard_2013,Tronko_Brizard_2015,Brizard_Tronci_2016} and we introduced the symmetric dyadic tensor
\begin{equation}
\mathbb{P}_{\|} \;=\; \frac{P_{\|}}{m\Omega}\;\left(\mathbb{I} - \bhat\bhat\right).
\label{eq:P_dyad}
\end{equation}
 In Eq.~\eqref{eq:deltaB_P}, and in what follows, we will use the dyadic double-dot product notation ${\bf a}\,{\bf b}\;\vb{:}\;\mathbb{W} = {\bf b}\bdot\mathbb{W}\bdot{\bf a} = b_{i}\,W^{ij}\,a_{j}$, where $({\bf a},{\bf b})$ is an arbitrary pair of vectors and $\mathbb{W}$ is an arbitrary second-rank tensor, and summation over repeated indices is assumed.

Next, the variations of the guiding-center kinetic energy are
 \begin{eqnarray}
 \delta_{\bf E}K_{\rm gc} &=& -\;\delta{\bf E}\bdot\frac{e\bhat}{\Omega}\btimes\left({\bf E}\btimes\frac{c\bhat}{B}\right) \;\equiv\; -\;\delta{\bf E}\bdot\vb{\pi}_{\rm E},  \label{eq:delta_E_K} \\
 \delta_{\bf B}K_{\rm gc} &=& \mu\,\bhat\bdot\delta{\bf B} \;-\; \delta{\bf B}\bdot\fd{}{\bf B}\left[ \frac{mc^{2}}{2\,B^{2}} {\bf E}\bdot\left(\mathbb{I} - \bhat\bhat\right)\bdot{\bf E}\right] \nonumber \\
  &\equiv&  -\;\delta{\bf B}\bdot\vb{\mu}_{\rm gc}.  \label{eq:delta_B_K}
\end{eqnarray}
where $\vb{\pi}_{\rm E}$ denotes the electric-dipole moment and the intrinsic guiding-center magnetic dipole is defined as
\begin{equation}
\vb{\mu}_{\rm gc} \;\equiv\; -\,\mu\,\bhat \;-\; \frac{\bf E}{B}\bdot\left(\bhat\,\vb{\pi}_{\rm E} + \vb{\pi}_{\rm E}\,\bhat\right),
\end{equation}
which includes the standard magnetization $-\mu\,\bhat$ as well as contributions from the electric-dipole moment $\vb{\pi}_{\rm E}$.

Using the definitions \eqref{eq:delta_E_Lag}-\eqref{eq:delta_B_Lag}, we, therefore, find
\begin{eqnarray}
\delta_{\bf E}L_{\rm gc} &=& \delta{\bf E}\bdot\vb{\pi}_{\rm E}, \\
\delta_{\bf B}L_{\rm gc} &=& \delta{\bf B}\bdot\left(\vb{\mu}_{\rm gc} + \vb{\pi}_{\rm gc}\btimes \frac{P_{\|}\bhat}{mc}\right),
\end{eqnarray}
so that the guiding-center polarization is defined as
\begin{equation}
\vb{\sf P}_{\rm gc} \;=\; \int_{P} F_{\rm gc}\,\vb{\pi}_{\rm E},
\label{eq:Pol_gc}
\end{equation}
while the guiding-center magnetization is defined as
\begin{equation}
\vb{\sf M}_{\rm gc} \;=\; \int_{P} F_{\rm gc}\left( \vb{\mu}_{\rm gc} + \vb{\pi}_{\rm gc}\times \frac{P_{\|}\bhat}{mc}\right).
  \label{eq:Mag_gc}
\end{equation}
Finally we note that the particle charge density $\varrho$ and the particle current density ${\bf J}$ are now expressed in guiding-center form as
\begin{eqnarray}
\varrho &=& \varrho_{\rm gc} \;-\; \nabla\bdot\vb{\sf P}_{\rm gc}, \label{eq:rho_particle} \\
{\bf J} &=& {\bf J}_{\rm gc} \;+\; \partial\vb{\sf P}_{\rm gc}/\partial t \;+\; c\,\nabla\btimes\vb{\sf M}_{\rm gc}, \label{eq:J_particle}
\end{eqnarray}
which guarantees the guiding-center charge conservation law \eqref{eq:charge_cons}.

\section{\label{sec:gc_Noether}Exact Energy-Momentum Conservation Laws}

Using the Noether method \citep{Brizard_2000_prl}, the guiding-center Vlasov-Maxwell equations \eqref{eq:gcVlasov_div}-\eqref{eq:Faraday} can be shown to possess exact energy-momentum conservation laws \citep{Brizard_2008}. We note that these conservation laws are exact since the guiding-center Vlasov-Maxwell Lagrangian and Hamiltonian are assumed independent of external electromagnetic fields. In what follows, we will also assume vanishing contributions at spatial boundaries, so that energy-momentum sources are omitted.

The guiding-center Vlasov-Maxwell energy conservation law 
\begin{equation}
\partial{\cal E}_{\rm gc}/\partial t \;+\; \nabla\bdot{\bf S}_{\rm gc} \;=\; 0 
\label{eq:energy_gc}
\end{equation}
is expressed in terms of the guiding-center Vlasov-Maxwell energy density
\begin{eqnarray}
{\cal E}_{\rm gc} &=& \int_{P} F_{\rm gc}\,K_{\rm gc} + \frac{\bf E}{4\pi}\bdot{\bf D}_{\rm gc} - \frac{1}{8\pi}\left(|{\bf E}|^{2} - |{\bf B}|^{2}\right) \nonumber \\
 &=& \int_{P} F_{\rm gc}\,K_{\rm gc} + {\bf E}\bdot\vb{\sf P}_{\rm gc} + \frac{1}{8\pi}\left(|{\bf E}|^{2} + |{\bf B}|^{2}\right),
 \label{eq:Egc_def}
\end{eqnarray}
and the guiding-center Vlasov-Maxwell energy density-flux
\begin{eqnarray}
{\bf S}_{\rm gc} &=& \int_{P} F_{\rm gc}\,\dot{\bf X}\,K_{\rm gc} \;+\; \frac{c\,{\bf E}}{4\pi}\btimes{\bf H}_{\rm gc}.
\label{eq:S_gc} 
\end{eqnarray}
 The guiding-center Vlasov-Maxwell energy density ${\cal E}_{\rm gc}$ will be used in Eq.~\eqref{eq:Hgc_def} to define the guiding-center Hamiltonian functional ${\cal H}_{\rm gc} \equiv \int_{X}{\cal E}_{\rm gc}$.

The guiding-center Vlasov-Maxwell momentum conservation law 
\begin{equation}
\partial{\bf P}_{\rm gc}/\partial t \;+\; \nabla\bdot\mathbb{T}_{\rm gc} \;=\; 0
\label{eq:momentum_gc} 
\end{equation}
is expressed in terms of the guiding-center Vlasov-Maxwell momentum density
\begin{equation}
{\bf P}_{\rm gc} \;=\; \int_{P} F_{\rm gc}\,P_{\|}\,\bhat \;+\; {\bf D}_{\rm gc}\btimes \frac{\bf B}{4\pi c},
\label{eq:Pgc_def}
\end{equation}
and the guiding-center Vlasov-Maxwell stress tensor is
\begin{eqnarray}
\mathbb{T}_{\rm gc} &\equiv& \int_{P} F_{\rm gc}\,\dot{\bf X}\,P_{\|}\,\bhat - \frac{1}{4\pi}\left({\bf D}_{\rm gc}\,{\bf E} + {\bf B}\,{\bf H}_{\rm gc}\right) \nonumber \\
 &&+\; \frac{\mathbb{I}}{4\pi} \left[ \frac{1}{2}\left(|{\bf E}|^{2} - |{\bf B}|^{2}\right) + {\bf B}\bdot{\bf H}_{\rm gc}\right],
 \label{eq:T_gc_tensor}
\end{eqnarray}
which can be written in symmetric dyadic form as
\begin{widetext}
\begin{eqnarray}
\mathbb{T}_{\rm gc} &=& \int_{P} F_{\rm gc} \left[ \frac{P_{\|}^{2}}{m}\;\bhat\,\bhat \;+\; \mu B\left(\mathbb{I} - \bhat\,\bhat\right) \;+\; P_{\|}\;\left(\bhat\;\dot{\bf X}_{\bot} + \dot{\bf X}_{\bot}\;\bhat\right)  \right] \nonumber \\
 &&+\;  \frac{1}{4\pi}\;\left[ \left(|{\bf E}|^{2} \;+\frac{}{} |{\bf B}^{2}|\right) \frac{\mathbb{I}}{2} \;-\; \left({\bf E}{\bf E} \;+\frac{}{} {\bf B}{\bf B} \right) \right] \;+\; \chi_{\rm gc} \left[|{\bf E}_{\bot}|^{2}\;\left(\mathbb{I} - \bhat\,\bhat\right) \;+\; E_{\|}^{2}\;\bhat\,\bhat \;-\; {\bf E}\,{\bf E}\right].
 \label{eq:T_gc_tensor_sym}
\end{eqnarray}
\end{widetext}
Here, we introduced the classical guiding-center susceptibility tensor $\overleftrightarrow{\vb{\chi}_{\rm gc}} \equiv \chi_{\rm gc}\,\left(\mathbb{I} - \bhat\,\bhat\right)$, where \citep{Northrop_1963,RGL_1981} 
\begin{equation}
\chi_{\rm gc}[F_{\rm gc},{\bf B}] \;\equiv\; \frac{mc^{2}}{B^{2}}\;\int_{P} F_{\rm gc} \;=\; \frac{c^{2}}{4\pi\,v_{\rm A}^{2}}
\label{eq:chi_def}
\end{equation}
is a functional of the magnetic field ${\bf B}$ as well as a functional of the guiding-center Vlasov phase-space density $F_{\rm gc}$.  Because of this symmetry property, we are also guaranteed that the guiding-center Vlasov-Maxwell toroidal angular momentum density
\begin{equation}
P_{{\rm gc}\varphi} \;\equiv\; \int_{P} F_{\rm gc}\,P_{\|}\,b_{\varphi} \;+\; {\bf D}_{\rm gc}\btimes \frac{\bf B}{4\pi c}\bdot\pd{\bf X}{\varphi}.
\label{eq:Pgc_varphi}
\end{equation}
satisfies the guiding-center Vlasov-Maxwell toroidal angular momentum conservation law
\begin{equation}
\pd{P_{{\rm gc}\varphi}}{t} \;+\; \nabla\bdot\left(\mathbb{T}_{\rm gc}\bdot\pd{\bf X}{\varphi}\right) \;=\; \mathbb{T}_{\rm gc}^{\top}\;\vb{:}\;\nabla\left(\pd{\bf X}{\varphi}\right) \;\equiv\; 0,
\label{eq:momentum_gc_varphi} 
\end{equation}
which follows from the dyadic symmetry of the guiding-center stress tensor \eqref{eq:T_gc_tensor_sym}, and the antisymmetry of the dyadic tensor $\nabla(\partial_{\varphi}{\bf X})$, i.e.,  the  dyadic tensor identity $({\bf V}{\bf W} + 
{\bf W}{\bf V})\;\vb{:}\;\nabla(\partial_{\varphi}{\bf X}) \equiv 0$ follows from the vector identity ${\bf W}\,{\bf V}\;\vb{:}\;\nabla(\partial_{\varphi}{\bf X}) = \wh{\sf z}\bdot({\bf V}\btimes{\bf W})$, which holds for arbitrary vectors $({\bf V},{\bf W})$.

The guiding-center Vlasov-Maxwell energy conservation law \eqref{eq:energy_gc} yields the guiding-center Vlasov-Maxwell Hamiltonian functional
\begin{widetext}
\begin{eqnarray}
{\cal H}_{\rm gc}[F_{\rm gc},{\bf D}_{\rm gc},{\bf B}] &=& \int_{Z} F_{\rm gc}\,K_{\rm gc} + \int_{X} \left[\frac{\bf E}{4\pi}\bdot{\bf D}_{\rm gc} - \frac{1}{8\pi}\left(|{\bf E}|^{2} - |{\bf B}|^{2}\right)\right] \nonumber \\
 &\equiv& \int_{Z} F_{\rm gc}\,\left(\mu B + \frac{P_{\|}^{2}}{2m}\right) \;+\; \int_{X} \left[ \frac{|{\bf B}|^{2}}{8\pi} \;+\; \frac{{\bf D}_{\rm gc}}{8\pi}\bdot\left(\overleftrightarrow{\vb{\varepsilon}_{\rm gc}}\right)^{-1}\bdot{\bf D}_{\rm gc}\right],
\label{eq:Hgc_def}
\end{eqnarray}
\end{widetext}
where the guiding-center permittivity tensor $ \overleftrightarrow{\vb{\varepsilon}_{\rm gc}}$ is defined through the relation
\begin{equation}
{\bf D}_{\rm gc} \;\equiv\;  \overleftrightarrow{\vb{\varepsilon}_{\rm gc}}\bdot{\bf E} \;=\; \left[\mathbb{I} \;+\; 4\pi\;\chi_{\rm gc}\,\left(\mathbb{I} - \bhat\bhat\right)\right]\bdot{\bf E},
\label{eq:DE_E}
\end{equation}
which is inverted 
\begin{equation}
{\bf E}[F_{\rm gc},{\bf D}_{\rm gc},{\bf B}] \;\equiv\; \overleftrightarrow{\vb{\varepsilon}_{\rm gc}}^{-1}[F_{\rm gc},{\bf B}]\bdot{\bf D}_{\rm gc}
\label{eq:E_inverse}
\end{equation} 
in order to obtain the guiding-center Hamiltonian functional \eqref{eq:Hgc_def}.

The guiding-center Vlasov-Maxwell momentum and angular momentum conservation laws \eqref{eq:momentum_gc} and \eqref{eq:momentum_gc_varphi}, on the other hand, yield the guiding-center Vlasov-Maxwell momentum and angular momentum functionals
\begin{eqnarray}
\vb{\cal P}_{\rm gc} &=& \int_{Z} F_{\rm gc}\; P_{\|}\,\bhat \;+\; \int_{X} \left(\frac{{\bf D}_{\rm gc}\btimes{\bf B}}{4\pi\;c}\right).
\label{eq:mom_func_def} \\
{\cal P}_{{\rm gc}\varphi} &=& \int_{Z} F_{\rm gc}\; P_{\|}\,b_{\varphi} \;+\; \int_{X} \left(\frac{{\bf D}_{\rm gc}\btimes{\bf B}}{4\pi\;c}\right)\bdot\pd{\bf X}{\varphi}.
\label{eq:ang_mom_func_def}
\end{eqnarray}
The three guiding-center functionals \eqref{eq:Hgc_def} and \eqref{eq:mom_func_def}-\eqref{eq:ang_mom_func_def} will play an important role in constructing the Hamiltonian formulation of the guiding-center Vlasov-Maxwell equations \eqref{eq:gcVlasov_div}, \eqref{eq:gc_Maxwell}, and \eqref{eq:Faraday}.

\vspace*{0.2in}

\section{\label{sec:Ham_gcVM}Hamiltonian Formulation of the Guiding-center Vlasov-Maxwell Equations}

We note that the conserved guiding-center functionals ${\cal I}_{\rm gc} = ({\cal H}_{\rm gc}, 
\vb{\cal P}_{\rm gc}, {\cal P}_{{\rm gc}\varphi})$, defined in Eqs.~\eqref{eq:Hgc_def} and \eqref{eq:mom_func_def}-\eqref{eq:ang_mom_func_def}, are assumed to satisfy the guiding-center Vlasov-Maxwell functional conservation law
\begin{equation}
[{\cal I}_{\rm gc},\;{\cal H}_{\rm gc}]_{\rm gc} \;=\; 0,
\label{eq:I_gc}
\end{equation}
While the conservation of ${\cal H}_{\rm gc}$ is automatically satisfied by the antisymmetry of the guiding-center Hamiltonian functional bracket $[\;,\;]_{\rm gc}$, the remaining conservation laws present constraints on the guiding-center Vlasov-Maxwell bracket. 

In addition, the guiding-center entropy functional
\begin{equation}
{\cal S}_{\rm gc}[F_{\rm gc},{\bf B}] \;\equiv\; -\; \int_{Z} F_{\rm gc}\;\ln\left(F_{\rm gc}/{\cal J}_{\rm gc}\right),
\label{eq:gc_entropy}
\end{equation}
must be a Casimir functional of the guiding-center Hamiltonian functional bracket $[\;,\;]_{\rm gc}$, i.e., $\left[{\cal S}_{\rm gc},\frac{}{} {\cal K}\right]_{\rm gc} = 0$ must vanish for all functionals ${\cal K}$.

\subsection{Bracket construction}

In order to construct the Hamiltonian structure for the guiding-center Vlasov-Maxwell equations \eqref{eq:gcVlasov_div}, \eqref{eq:gc_Maxwell}, and \eqref{eq:Faraday}, we now follow the same procedure used to construct the Hamiltonian functional bracket $[\;,\;]$ for the Vlasov-Maxwell equations (reviewed in App.~\ref{sec:VM}), and we consider the time evolution of an arbitrary functional 
${\cal F}[F_{\rm gc},{\bf D}_{\rm gc},{\bf B}]$:
\begin{eqnarray}
\pd{\cal F}{t} &=& \int_{Z}\pd{F_{\rm gc}}{t}\,\fd{\cal F}{F_{\rm gc}} + \int_{X} \pd{{\bf D}_{\rm gc}}{t}\bdot\fd{\cal F}{{\bf D}_{\rm gc}} + \pd{\bf B}{t}\bdot\fd{\cal F}{\bf B} 
\nonumber  \\
 &=& \int_{Z} F_{\rm gc} \left[ \dot{Z}^{\alpha}\,\pd{}{Z^{\alpha}}\left(\fd{\cal F}{F_{\rm gc}}\right) - 4\pi e\,\dot{\bf X}\bdot\fd{\cal F}{{\bf D}_{\rm gc}}\right] \nonumber \\
  &&+\; c\int_{X} \left(\fd{\cal F}{{\bf D}_{\rm gc}}\bdot\nabla\btimes{\bf H}_{\rm gc} - \fd{\cal F}{\bf B}\bdot\nabla\btimes{\bf E} \right),
  \label{eq:partial_F}
\end{eqnarray}
after inserting the guiding-center Vlasov-Maxwell equations \eqref{eq:gcVlasov_div}, \eqref{eq:gc_Maxwell}, and \eqref{eq:Faraday}. 

Using the Hamiltonian functional \eqref{eq:Hgc_def} explicitly as a functional of the guiding-center Vlasov-Maxwell fields $(F_{\rm gc},{\bf D}_{\rm gc},{\bf B})$, and using the matrix identity $\delta\left(\overleftrightarrow{\vb{\varepsilon}_{\rm gc}}\right)^{-1} \equiv 
-\; \left(\overleftrightarrow{\vb{\varepsilon}_{\rm gc}}\right)^{-1}\bdot\delta\left(\overleftrightarrow{\vb{\varepsilon}_{\rm gc}}\right)\bdot\left(\overleftrightarrow{\vb{\varepsilon}_{\rm gc}}\right)^{-1}$, we find the Hamiltonian functional derivatives
\begin{eqnarray}
\fd{{\cal H}_{\rm gc}}{F_{\rm gc}} &=& \mu\,B + \frac{P_{\|}^{2}}{2m} + \frac{{\bf D}_{\rm gc}}{8\pi}\bdot \fd{\left(\overleftrightarrow{\vb{\varepsilon}_{\rm gc}}\right)^{-1}}{F_{\rm gc}}\bdot{\bf D}_{\rm gc} \nonumber \\
 &=& \mu\,B + \frac{P_{\|}^{2}}{2m} - \frac{\bf E}{8\pi}\bdot \fd{\overleftrightarrow{\vb{\varepsilon}_{\rm gc}}}{F_{\rm gc}}\bdot{\bf E} \nonumber \\
  &=&  \mu\,B + \frac{P_{\|}^{2}}{2m} - \frac{mc^{2}}{2B^{2}}\;{\bf E}\bdot\left(\mathbb{I} - \bhat\bhat\right)\bdot{\bf E} \nonumber \\
 &=&\mu\,B + \frac{P_{\|}^{2}}{2m} - \frac{m}{2}\,|{\bf u}_{\rm E}|^{2} = K_{\rm gc}, \label{eq:H_F} \\
\fd{{\cal H}_{\rm gc}}{{\bf D}_{\rm gc}} &=& \left(\overleftrightarrow{\vb{\varepsilon}_{\rm gc}}\right)^{-1}\bdot{\bf D}_{\rm gc}/4\pi \;=\; {\bf E}/4\pi, \label{eq:H_D} \\
\fd{{\cal H}_{\rm gc}}{\bf B} &=& \frac{\bf B}{4\pi} + \fd{}{\bf B}\left[\frac{{\bf D}_{\rm gc}}{8\pi}\bdot\left(\overleftrightarrow{\vb{\varepsilon}_{\rm gc}}\right)^{-1}\bdot{\bf D}_{\rm gc}\right] + \int_{P} F_{\rm gc}\;\mu\,\bhat \nonumber \\
 &=& \frac{\bf B}{4\pi} \;-\; \int_{P} F_{\rm gc}\;\vb{\mu}_{\rm gc}, \label{eq:H_B}
\end{eqnarray}
where we used the identity \eqref{eq:E_inverse} in obtaining Eq.~\eqref{eq:H_F}. In Eq.~\eqref{eq:H_B}, the intrinsic magnetic-dipole moment $\vb{\mu}_{\rm gc}$ is defined in Eq.~\eqref{eq:delta_B_K}, which yields the relation
\begin{equation}
\frac{{\bf H}_{\rm gc}}{4\pi} \;=\; \fd{{\cal H}_{\rm gc}}{\bf B} \;-\; \int_{P} F_{\rm gc}\;\left(\mathbb{P}_{\|}\bdot\frac{e}{c}\;\dot{\bf X}\right),
\end{equation}
where we used the definition \eqref{eq:Mag_gc}. Hence, the last line of Eq.~\eqref{eq:partial_F} is expressed as
\begin{eqnarray}
&&4\pi\,c \int_{X} \left( \fd{\cal F}{{\bf D}_{\rm gc}}\bdot\nabla\btimes\fd{{\cal H}_{\rm gc}}{\bf B} \;-\; \fd{\cal F}{\bf B}\bdot\nabla\btimes\fd{{\cal H}_{\rm gc}}{{\bf D}_{\rm gc}}\right) \nonumber \\
 &&-\; 4\pi\,e\int_{Z} \fd{\cal F}{{\bf D}_{\rm gc}}\bdot\nabla\btimes\left(F_{\rm gc}\;\mathbb{P}_{\|}\bdot\frac{e}{c}\;\dot{\bf X} \right),
\label{eq:partial_F_2}
\end{eqnarray}
while the remaining terms are expressed as
\begin{eqnarray}
\dot{Z}^{\alpha} &=& \left\{Z^{\alpha},\; \fd{{\cal H}_{\rm gc}}{F_{\rm gc}}\right\}_{\rm gc} + 4\pi e\,\frac{\delta^{\star}{\cal H}_{\rm gc}}{\delta{\bf D}_{\rm gc}}\bdot\{{\bf X}, Z^{\alpha}\}_{\rm gc}, \label{eq:Z_dot_separate} \\
\dot{\bf X} &=& \left\{{\bf X},\;  \fd{{\cal H}_{\rm gc}}{F_{\rm gc}}\right\}_{\rm gc} + 4\pi e\,\frac{\delta^{\star}{\cal H}_{\rm gc}}{\delta{\bf D}_{\rm gc}}\bdot\{{\bf X}, {\bf X}\}_{\rm gc}, \label{eq:X_dot_separate} 
\end{eqnarray}
where, using Faraday's law \eqref{eq:Faraday}, we introduced the definition of an effective functional derivative $\delta^{\star}(\;)/\delta{\bf D}_{\rm gc}$:
\begin{eqnarray}
e\,{\bf E}^{*}&=& e\,{\bf E} \;-\; \frac{P_{\|}}{B}\left(\mathbb{I} - \bhat\bhat\right)\bdot\pd{\bf B}{t} 
\label{eq:E_star_Ham} \\
 &=& e\,{\bf E} \;+\; \frac{c\,P_{\|}}{B}\left(\mathbb{I} - \bhat\bhat\right)\bdot\nabla\btimes{\bf E} \nonumber \\ 
 &=& 4\pi e \left(\fd{{\cal H}_{\rm gc}}{{\bf D}_{\rm gc}} \;+\; \mathbb{P}_{\|}\bdot\nabla\btimes\fd{{\cal H}_{\rm gc}}{{\bf D}_{\rm gc}}\right) \;\equiv\; 4\pi e\,\frac{\delta^{\star}{\cal H}_{\rm gc}}{\delta{\bf D}_{\rm gc}}.
\nonumber
\end{eqnarray}
By integrating by parts the last term in Eq.~\eqref{eq:partial_F_2}, and inserting the definition of the effective functional derivative 
$\delta^{\star}(\;)/\delta{\bf D}_{\rm gc}$, we obtain
\[ \int_{Z} \fd{\cal F}{{\bf D}_{\rm gc}}\bdot\nabla\btimes\left(F_{\rm gc}\;\mathbb{P}_{\|}\bdot\frac{e}{c}\;\dot{\bf X} \right) = \int_{Z} F_{\rm gc}\,\frac{\delta^{\star}{\cal F}}{\delta{\bf D}_{\rm gc}}\bdot\frac{e}{c}\;\dot{\bf X}, \]
where $\dot{\bf X}$ is given by Eq.~\eqref{eq:X_dot_separate}.

By combining these expressions, with summation over particle species implied wherever appropriate, we find the final expression for Eq.~\eqref{eq:partial_F}
\begin{widetext}
 \begin{eqnarray}
\pd{\cal F}{t} &=& \int_{Z} F_{\rm gc} \left\{ \fd{{\cal F}}{F_{\rm gc}} ,\; \fd{{\cal H}_{\rm gc}}{F_{\rm gc}} \right\}_{\rm gc} + 4\pi e \int_{Z} F_{\rm gc}\left( \frac{\delta^{\star}{\cal H}_{\rm gc}}{\delta{\bf D}_{\rm gc}}\vb{\cdot}\left\{{\bf X}, \fd{{\cal F}}{F_{\rm gc}} \right\}_{\rm gc} - \frac{\delta^{\star}{\cal F}}{\delta{\bf D}_{\rm gc}}\vb{\cdot}\left\{{\bf X}, \fd{{\cal H}_{\rm gc}}{F_{\rm gc}} 
 \right\}_{\rm gc} \right) \nonumber \\
   &&+\;  (4\pi e)^{2} \int_{\bf Z} F_{\rm gc} \left(\frac{\delta^{\star}{\cal F}}{\delta{\bf D}_{\rm gc}}\bdot\left\{{\bf X}, {\bf X}\right\}_{\rm gc}\bdot \frac{\delta^{\star}{\cal H}_{\rm gc}}{\delta{\bf D}_{\rm gc}}\right) \;+\; 4\pi c  \int_{X} 
   \left(\fd{\cal F}{\bf D}_{\rm gc}\vb{\cdot}\nabla\vb{\times}\fd{{\cal H}_{\rm gc}}{\bf B} \;-\; \fd{{\cal H}_{\rm gc}}{\bf D}_{\rm gc}\vb{\cdot}\nabla\vb{\times}\fd{\cal F}{\bf B} \right).
  \label{eq:F_H}
 \end{eqnarray}
from which Eq.~\eqref{eq:F_HB} yields the guiding-center Vlasov-Maxwell bracket
 \begin{eqnarray}
 \left[{\cal F},\frac{}{}{\cal G}\right]_{\rm gc} &=& \int_{Z} F_{\rm gc} \left\{ \fd{{\cal F}}{F_{\rm gc}} , \fd{\cal G}{F_{\rm gc}} \right\}_{\rm gc} \;+\; 4\pi e \int_{Z} F_{\rm gc}\;\left( \frac{\delta^{\star}{\cal G}}{\delta{\bf D}_{\rm gc}}\vb{\cdot}
 \left\{{\bf X},\; \fd{{\cal F}}{F_{\rm gc}} \right\}_{\rm gc} - \frac{\delta^{\star}{\cal F}}{\delta{\bf D}_{\rm gc}}\vb{\cdot}\left\{{\bf X},\;\fd{\cal G}{F_{\rm gc}} \right\}_{\rm gc} 
  \right) \nonumber \\
   &&+\; (4\pi e)^{2} \int_{\bf Z} F_{\rm gc} \left(\frac{\delta^{\star}{\cal F}}{\delta{\bf D}_{\rm gc}}\bdot\left\{{\bf X},\; {\bf X}\right\}_{\rm gc}\bdot \frac{\delta^{\star}{\cal G}}{\delta{\bf D}_{\rm gc}}\right) \;+\; 4\pi c  \int_{X} 
   \left(\fd{\cal F}{\bf D}_{\rm gc}\vb{\cdot}\nabla\vb{\times}\fd{\cal G}{\bf B} \;-\; \fd{\cal G}{\bf D}_{\rm gc}\vb{\cdot}\nabla\vb{\times}\fd{\cal F}{\bf B} \right),
  \label{eq:gcVM_bracket}
 \end{eqnarray}
 \end{widetext}
 where ${\cal F}$ and ${\cal G}$ are arbitrary functionals of the guiding-center Vlasov-Maxwell fields $(F_{\rm gc},{\bf D}_{\rm gc},{\bf B})$. We note that the antisymmetry, $[{\cal F},\;{\cal G}]_{\rm gc} = -\,[{\cal G},\;{\cal F}]_{\rm gc}$, of the guiding-center Vlasov-Maxwell bracket \eqref{eq:gcVM_bracket} depends crucially on the separation \eqref{eq:Xgc_dot}-\eqref{eq:Pgc_dot} [see also Eq.~\eqref{eq:Z_dot_separate}] between contributions from the guiding-center kinetic energy $K_{\rm gc} \equiv \delta{\cal H}_{\rm gc}/\delta F_{\rm gc}$ and the modified electric field ${\bf E}^{*} \equiv 4\pi\,\delta^{\star}{\cal H}_{\rm gc}/\delta{\bf D}_{\rm gc}$ to the guiding-center Hamiltonian dynamics in reduced guiding-center phase space $({\bf X},P_{\|})$. The same separation of the particle equations of motion was responsible for the antisymmetric bracket derived by Morrison \cite{Morrison_1980,Morrison_1982} and Marsden and Weinstein \cite{Marsden_Weinstein_1982} for the Vlasov-Maxwell equations (see Eqs.~\eqref{eq:z_dot} and \eqref{eq:f_H} in App.~\ref{sec:VM}).

The guiding-center Vlasov-Maxwell bracket \eqref{eq:gcVM_bracket} is identical in form to the guiding-center bracket previously derived for the guiding-center Vlasov-Maxwell equations without polarization \citep{Brizard_2021_gcVM}, except that functional derivatives $(\delta/\delta{\bf E}, \delta^{\star}/\delta{\bf E})$ with respect to the electric field ${\bf E}$ are now replaced with functional derivatives $(\delta/\delta{\bf D}_{\rm gc}, \delta^{\star}/\delta{\bf D}_{\rm gc})$ with respect to the guiding-center displacement field ${\bf D}_{\rm gc}$. Once again, this replacement is caused by the fact that, in the noncanonical formulation of the guiding-center Vlasov-Maxwell equations adopted here, the guiding-center magnetization $\vb{\sf M}_{\rm gc}$ is no longer uniquely defined by the functional derivative $\partial K_{\rm gc}/\partial{\bf B}$ of the guiding-center kinetic energy [see Eqs.~\eqref{eq:delta_B_K} and \eqref{eq:Mag_gc}], as would be the case in the canonical formulation adopted in Ref.~\cite{Pfirsch_Morrison_1985} and assumed in Refs.~\cite{Morrison_2013} and \cite{Burby_BMQ_2015}.

\subsection{Jacobi property}

The proof of the Jacobi property
 \begin{eqnarray}
 0 &=& {\cal Jac}[{\cal F},{\cal G},{\cal K}] 
  \label{eq:Jac_gcVM} \\
  &\equiv& \left[{\cal F},\frac{}{}[{\cal G}, {\cal K}]_{\rm gc}\right]_{\rm gc} +   \left[{\cal G},\frac{}{}[{\cal K}, {\cal F}]_{\rm gc}\right]_{\rm gc}  +   \left[{\cal K},\frac{}{}[{\cal F}, {\cal G}]_{\rm gc}\right]_{\rm gc} 
\nonumber
 \end{eqnarray}
 for the guiding-center Vlasov-Maxwell bracket \eqref{eq:gcVM_bracket}, which must hold for any guiding-center functionals $({\cal F},{\cal G},{\cal K})$, is, therefore, identical to the proof presented in Ref.~\cite{Brizard_gcVM_proof}. 
 
 The Jacobi property \eqref{eq:Jac_gcVM}, which is inherited from the Jacobi property \eqref{eq:Jac_gcPB}, can thus be expressed as a series in powers of $(4\pi e)$ up to third order \citep{Brizard_gcVM_proof}:
\begin{widetext}
\begin{eqnarray}
{\cal Jac}[{\cal F},{\cal G},{\cal K}]  &=& \int_{Z} F_{\rm gc}\;{\sf Jac}[f,g,k] \;+\; 4\pi e\; \int_{Z} F_{\rm gc}\; \left( F_{i}^{\star}\; {\sf Jac}[X^{i};g,k] + \leftturn \right)    \label{eq:Jacobi_4piq}  \\
  &&+ (4\pi e)^{2}\; \int_{Z} F_{\rm gc} \;\left( F_{i}^{\star}\,G_{j}^{\star}\; {\sf Jac}[X^{i},X^{j};k] + \leftturn \right) \;+\; (4\pi e)^{3}\; \int_{Z} F_{\rm gc} \;\left(F_{i}^{\star}\,G_{j}^{\star}\,K_{\ell}^{\star}\frac{}{}{\sf Jac}[X^{i},X^{j},X^{\ell}]\right),
\nonumber
\end{eqnarray}
\end{widetext}
where we use the notation $(f,g,k) \equiv (\delta{\cal F}/\delta F_{\rm gc}, \delta{\cal G}/\delta F_{\rm gc}, \delta{\cal K}/\delta F_{\rm gc})$ and $({\bf F}^{\star},{\bf G}^{\star},{\bf K}^{\star}) \equiv (\delta^{\star}{\cal F}/\delta {\bf D}_{\rm gc}, \delta^{\star}{\cal G}/\delta{\bf D}_{\rm gc}, \delta^{\star}{\cal K}/\delta{\bf D}_{\rm gc})$, the guiding-center Jacobiator ${\sf Jac}[\;,\;,\;]$ is defined in Eq.~\eqref{eq:Jac_gcPB}. In addition, the $i^{th}$-component of the guiding-center position 
${\bf X}$ is denoted as $X^{i}$, summation over repeated indices is assumed, and the symbol $\leftturn$ denotes a cyclic permutation of the functionals $({\cal F},{\cal G},{\cal K})$. In Eq.~\eqref{eq:Jacobi_4piq}, it is, therefore, clear that the Jacobi property of the guiding-center Vlasov-Maxwell bracket \eqref{eq:gcVM_bracket} is inherited from the Jacobi property \eqref{eq:Jac_gcPB} of the guiding-center Poisson bracket \eqref{eq:gcPB}, since each term in Eq.~\eqref{eq:Jacobi_4piq} vanishes identically because of this latter property, i.e., ${\sf Jac}[F,G,K] = 0$ for all functions $(F,G,K)$. Hence, the Jacobi property for the guiding-center Vlasov-Maxwell bracket \eqref{eq:gcVM_bracket} holds under the condition $\nabla\bdot{\bf B}^{*} = 0$.

 With this same replacement $({\bf E} \rightarrow {\bf D}_{\rm gc})$, we can easily see that the guiding-center Vlasov-Maxwell momentum functional \eqref{eq:mom_func_def} is a Hamiltonian functional invariant, while the guiding-center Vlasov-Maxwell entropy functional \eqref{eq:gc_entropy} is a Casimir invariant of the guiding-center Vlasov-Maxwell bracket \eqref{eq:gcVM_bracket}, with the explicit proofs presented in \cite{Brizard_2021_gcVM}.

\subsection{Hamiltonian conservation laws}

The energy-momentum and angular-momentum conservation laws of the guiding-center Vlasov-Maxwell equations can be expressed in Hamiltonian form \eqref{eq:I_gc}. In this Section, we show that the guiding-center angular momentum functional \eqref{eq:ang_mom_func_def} satisfies the guiding-center Hamiltonian conservation law \eqref{eq:I_gc}.

First, we rearrange terms in Eq.~\eqref{eq:gcVM_bracket} and we evaluate the functional derivatives of the guiding-center angular momentum functional \eqref{eq:ang_mom_func_def} 
\begin{eqnarray}
\fd{{\cal P}_{{\rm gc}\varphi}}{F_{\rm gc}} &=& P_{\|}b_{\varphi}, \\
4\pi c\,\fd{{\cal P}_{{\rm gc}\varphi}}{{\bf D}_{\rm gc}} &=& {\bf B}\btimes\pd{\bf X}{\varphi}, \\
4\pi c\,\fd{{\cal P}_{{\rm gc}\varphi}}{\bf B} &=& 4\pi\,e\int_{P} F_{\rm gc}\; \mathbb{P}_{\|}\bdot\pd{\bf X}{\varphi} + \pd{\bf X}{\varphi}\btimes{\bf D}_{\rm gc},
\end{eqnarray} 
which yields the guiding-center Vlasov-Maxwell bracket expression
 \begin{widetext}
 \begin{eqnarray}
 \left[{\cal P}_{{\rm gc}\varphi},\frac{}{}{\cal H}_{\rm gc}\right]_{\rm gc} &=& \int_{Z} F_{\rm gc} \left( \left\{ \fd{{\cal P}_{{\rm gc}\varphi}}{F_{\rm gc}},\; \fd{{\cal H}_{\rm gc}}{F_{\rm gc}} \right\}_{\rm gc} \;-\; 4\pi\,e\;
 \frac{\delta^{\star}{\cal P}_{{\rm gc}\varphi}}{\delta{\bf D}_{\rm gc}}\bdot\left\{{\bf X},\;\fd{{\cal H}_{\rm gc}}{F_{\rm gc}} \right\}_{\rm gc} \right) \nonumber \\
  &&+\; 4\pi e \int_{Z} F_{\rm gc}\;\left( 4\pi\,e\;\frac{\delta^{\star}{\cal P}_{{\rm gc}\varphi}}{\delta{\bf D}_{\rm gc}}\vb{\cdot}\left\{{\bf X}, {\bf X}\right\}_{\rm gc} \;+\; \left\{{\bf X},\; \fd{{\cal P}_{{\rm gc}\varphi}}{F_{\rm gc}} \right\}_{\rm gc} \right) \bdot\frac{\delta^{\star}{\cal H}_{\rm gc}}{\delta{\bf D}_{\rm gc}} \nonumber \\
    &&+ 4\pi c  \int_{X} \left(\fd{{\cal H}_{\rm gc}}{\bf B}\bdot\nabla\btimes\fd{{\cal P}_{{\rm gc}\varphi}}{{\bf D}_{\rm gc}} - 
  \fd{{\cal P}_{{\rm gc}\varphi}}{\bf B}\bdot\nabla\btimes\fd{{\cal H}_{\rm gc}}{{\bf D}_{\rm gc}} \right),
  \label{eq:Pgc_varphi_cons}
 \end{eqnarray}
\end{widetext}
Next, using the definition $\delta^{\star}(\cdots)/\delta{\bf D}_{\rm gc}$ given in Eq.~\eqref{eq:E_star_Ham}, we find the expression
\begin{equation}
4\pi\,e\;\frac{\delta^{\star}{\cal P}_{{\rm gc}\varphi}}{\delta{\bf D}_{\rm gc}} \;=\; \frac{e}{c}\,{\bf B}^{*}\btimes\pd{\bf X}{\varphi} \;+\; P_{\|}\,\nabla b_{\varphi},
\end{equation}
which can be used to obtain the following identities
\begin{eqnarray*}
4\pi\,e\;\frac{\delta^{\star}{\cal P}_{{\rm gc}\varphi}}{\delta{\bf D}_{\rm gc}}\bdot\left\{{\bf X},\fd{{\cal H}_{\rm gc}}{F_{\rm gc}} \right\}_{\rm gc}  &\equiv& \left\{ \fd{{\cal P}_{{\rm gc}\varphi}}{F_{\rm gc}} , \fd{{\cal H}_{\rm gc}}{F_{\rm gc}} 
\right\}_{\rm gc}  \\
 &&+\; \pd{\bf X}{\varphi}\bdot\nabla K_{\rm gc}, \\
4\pi\,e\;\frac{\delta^{\star}{\cal P}_{{\rm gc}\varphi}}{\delta{\bf D}_{\rm gc}}\bdot\left\{{\bf X},{\bf X}\right\}_{\rm gc}  &\equiv& \pd{\bf X}{\varphi} - \left\{{\bf X},\; \fd{{\cal P}_{{\rm gc}\varphi}}{F_{\rm gc}} \right\}_{\rm gc},
\end{eqnarray*}
so that Eq.~\eqref{eq:Pgc_varphi_cons} becomes
\begin{widetext}
 \begin{eqnarray}
 \left[{\cal P}_{{\rm gc}\varphi},\frac{}{}{\cal H}_{\rm gc}\right]_{\rm gc} &=& \int_{Z} F_{\rm gc} \left( e\,{\bf E}^{*} -\frac{}{} \nabla K_{\rm gc} \right)\bdot\pd{\bf X}{\varphi} \;+\;  4\pi c  \int_{X} \left( \fd{{\cal H}_{\rm gc}}{\bf B}\bdot\nabla\btimes\fd{{\cal P}_{{\rm gc}\varphi}}{{\bf D}_{\rm gc}} \;-\; \fd{{\cal P}_{{\rm gc}\varphi}}{\bf B}\bdot\nabla\btimes\fd{{\cal H}_{\rm gc}}{{\bf D}_{\rm gc}}\right).
 \label{eq:Pgc_varphi_final}
\end{eqnarray}
Next, we evaluate the remaining terms in Eq.~\eqref{eq:Pgc_varphi_final}:
\begin{eqnarray*}
4\pi c\;  \fd{{\cal H}_{\rm gc}}{\bf B}\bdot\nabla\btimes\fd{{\cal P}_{{\rm gc}\varphi}}{{\bf D}_{\rm gc}} &=& \left(\frac{\bf B}{4\pi} - \int_{P}F_{\rm gc}\,\vb{\mu}_{\rm gc}\right)\bdot\nabla\btimes\left({\bf B}\vb{\times}
\pd{\bf X}{\varphi}\right) \;=\; \left(\frac{\bf B}{4\pi} - \int_{P}F_{\rm gc}\,\vb{\mu}_{\rm gc}\right)\bdot\left(\pd{\bf B}{\varphi} \;+\; \wh{\sf z}\btimes{\bf B}\right) \\
 &=& \int_{P} F_{\rm gc} \left[\pd{\bf X}{\varphi}\bdot\nabla K_{\rm gc} + \left(\pd{\bf E}{\varphi} + \wh{\sf z}\btimes{\bf E}\right)\bdot\vb{\pi}_{E} \right] \;+\; \nabla\bdot\left(\pd{\bf X}{\varphi}\;\frac{|{\bf B}|^{2}}{8\pi}\right), \\
-\,4\pi c\;  \fd{{\cal P}_{{\rm gc}\varphi}}{\bf B}\cdot\nabla\vb{\times}\fd{{\cal H}_{\rm gc}}{{\bf D}_{\rm gc}} &=& -\,\left(\pd{\bf X}{\varphi}\btimes{\bf D}_{\rm gc}\right)\bdot\nabla\btimes\left(\frac{\bf E}{4\pi}\right) \;+\; \int_{P}F_{\rm gc}\;
e\,\pd{\bf X}{\varphi}\bdot\left({\bf E} \;-\frac{}{} {\bf E}^{*}\right) \\
 &=& -\,\int_{P} F_{\rm gc} \left[ \left(\pd{\bf E}{\varphi} + \wh{\sf z}\vb{\times}{\bf E}\right)\bdot\vb{\pi}_{E} \;+\; e\,{\bf E}^{*}\bdot\pd{\bf X}{\varphi} \right] + \nabla\bdot\left[ \frac{{\bf D}_{\rm gc}}{4\pi}\;{\bf E}\bdot\pd{\bf X}{\varphi} \;-\; \pd{\bf X}{\varphi}\;\left(\frac{|{\bf E}|^{2}}{8\pi}\right)\right],
\end{eqnarray*}
\end{widetext}
where we used the definitions \eqref{eq:E_star_Ham} and
\[ \pd{\bf X}{\varphi}\bdot\nabla K_{\rm gc} \;=\; \pd{K_{\rm gc}}{\varphi} \;\equiv\; -\,\pd{\bf E}{\varphi}\bdot\vb{\pi}_{\rm E} \;-\; \pd{\bf B}{\varphi}\bdot\vb{\mu}_{\rm gc}, \]
which follows from Eqs.~\eqref{eq:delta_E_K}-\eqref{eq:delta_B_K}. Hence, after omitting exact spatial divergences, Eq.~\eqref{eq:Pgc_varphi_final} becomes the guiding-center Hamiltonian conservation law
\begin{equation}
 \left[{\cal P}_{{\rm gc}\varphi},\frac{}{}{\cal H}_{\rm gc}\right]_{\rm gc}  \;\equiv\; 0,
\end{equation}
which proves that the guiding-center angular momentum functional \eqref{eq:ang_mom_func_def} is a functional invariant.

\section{Summary and Future Work}

In the present work, we showed that the guiding-center Vlasov-Maxwell equations \eqref{eq:gcVlasov_div}-\eqref{eq:Faraday}, which were previously derived from a variational principle \cite{Brizard_2021_gyVM}, can be expressed in Hamiltonian form \eqref{eq:F_HB} in terms of the Hamiltonian functional \eqref{eq:Hgc_def} and the guiding-center Vlasov-Maxwell bracket \eqref{eq:gcVM_bracket}. The present guiding-center Vlasov-Maxwell Hamiltonian structure generalizes our previous work  \cite{Brizard_2021_gcVM}, which allows a more consistent treatment of guiding-center polarization and magnetization. 

We remind the Reader that we adopted a {\it Hamiltonian} polarization representation, here, in which the guiding-center polarization is introduced explicitly through the guiding-center Hamiltonian. In the alternate {\it symplectic} polarization representation \cite{Brizard_2024}, the guiding-center polarization effects appear in both the symplectic (Poisson-bracket) structure as well as the guiding-center Hamiltonian, which implies that the polarization drift now appears as a guiding-center drift. This fact increases the complexity of the task of finding the Hamiltonian structure of the guiding-center Vlasov-Maxwell equations in the {\it symplectic} polarization representation, which will be left for future work.

In future work, we will also extend the work on the metriplectic foundations of gyrokinetic Vlasov–Maxwell–Landau theory \cite{Hirvijoki_2022} to the metriplectic foundations of guiding-center Vlasov–Maxwell–Landau theory to explore neoclassical transport in general magnetic geometry. 

\vspace*{0.2in}

\acknowledgments

The present work was supported by the National Science Foundation grant PHY-2206302.

\vspace*{0.1in}

\noindent
{\bf AUTHOR DECLARATIONS}

\vspace*{0.1in}

\noindent
{\bf Conflict of Interest} 

\vspace*{0.1in}

The author has no conflicts to disclose.

\appendix

\section{\label{sec:VM}Vlasov-Maxwell Hamiltonian Theory}

In this Appendix, the Hamiltonian structure for the Vlasov-Maxwell equations, which was originally discovered by  Morrison \cite{Morrison_1982} \&  Marsden and Weinstein \cite{Marsden_Weinstein_1982}, is presented using a direct construction that will be used in Sec.~\ref{sec:Ham_gcVM} to construct the Hamiltonian structure of the guiding-center Vlasov-Maxwell equations. In particular, we will show that the Hamiltonian structure for the Vlasov-Maxwell equations cannot be based on an extended Hamiltonian description of charged-particle dynamics in eight-dimensional phase space $({\bf x},{\bf p};t,w)$ that includes the time-energy canonically-conjugate coordinates $(t,w)$.

\subsection{Hamiltonian Particle Motion in Extended Phase Space} 

The equations of motion for a charged particle (of mass $m$ and charge $e$) can be expressed in Hamiltonian form 
\begin{equation}
dz^{a}/d\tau \;=\; \{ z^{a}, H_{\rm e}\}_{\rm e}
\label{eq:Hame_eq}
\end{equation}
 in extended phase space $z^{a} = ({\bf x},{\bf p}; t,w)$, where ${\bf p}$ denotes the kinetic momentum, $(t,w)$ denote the time-energy canonical coordinates, and $\tau$ is the Hamiltonian orbit parameter. Here, the extended Hamiltonian is defined as 
\begin{equation}
H_{\rm e} \;\equiv\; K \;+\; e\,\Phi \;-\; w, 
\label{eq:Hame_def}
\end{equation}
where $K \equiv |{\bf p}|^{2}/2m$ denotes the kinetic energy, and the extended (noncanonical) Poisson bracket is defined as
\begin{eqnarray} 
\{ F,\; G\}_{\rm e} &\equiv& \left(\pd{F}{w}\,\pd{G}{t} - \pd{F}{t}\,\pd{G}{w}\right) \nonumber \\
 &&+\; \left(\nabla_{\rm e}F\bdot\pd{G}{\bf p} \;-\; \pd{F}{\bf p}\bdot\nabla_{\rm e}G\right) \nonumber \\
 &&+\; \frac{e}{c}\,{\bf B}\bdot\left(\pd{F}{\bf p}\btimes\pd{G}{\bf p}\right),
\label{eq:PB_e}
\end{eqnarray}
where $(F,G)$ are arbitrary functions on extended phase space $({\bf x},{\bf p}; t,w)$, and  the extended gradient $\nabla_{\rm e}$ is defined as 
\[ \nabla_{\rm e} \;\equiv\; \nabla \;-\; \frac{e}{c}\,\pd{\bf A}{t}\;\pd{}{w}, \]
so that, using Eq.~\eqref{eq:Hame_def}, we obtain $\nabla_{\rm e}H_{\rm e} \equiv -e\,{\bf E}$. 

The equations of motion \eqref{eq:Hame_eq} can thus be explicitly expressed in extended Hamiltonian form as
\begin{eqnarray}
\frac{d{\bf x}}{d\tau} &=& \{ {\bf x},\; H_{\rm e}\}_{\rm e} \;=\; \pd{H_{\rm e}}{\bf p} \;=\; \frac{\bf p}{m}, 
\label{eq:xe_dot} \\
\frac{d{\bf p}}{d\tau} &=& \{ {\bf p},\; H_{\rm e}\}_{\rm e} \;=\; -\,\nabla_{\rm e}H_{\rm e} \;+\; \pd{H_{\rm e}}{\bf p}\btimes\frac{e}{c}\,{\bf B} \nonumber \\
 &=& e\,{\bf E} \;+\; \frac{e}{c}\,\frac{d{\bf x}}{d\tau}\btimes{\bf B}, 
\label{eq:pe_dot} \\
\frac{dt}{d\tau} &=& \{ t,\; H_{\rm e}\}_{\rm e} \;=\; -\;\pd{H_{\rm e}}{w} \;=\; 1,
\label{eq:te_dot}  \\
\frac{dw}{d\tau} &=&  \{ w,\; H_{\rm e}\}_{\rm e} \;=\; \pd{H_{\rm e}}{t} \;-\; \frac{e}{c}\,\pd{\bf A}{t}\bdot\pd{H_{\rm e}}{\bf p} \nonumber \\
 &=& e\,\pd{\Phi}{t} \;-\; \frac{e}{c}\,\pd{\bf A}{t}\bdot\frac{d{\bf x}}{d\tau},
\label{eq:we_dot} 
\end{eqnarray}
where, according to Eq.~\eqref{eq:te_dot}, the Hamiltonian orbit parameter $\tau$ is simply related to the time coordinate $t$. In addition, Eq.~\eqref{eq:we_dot} implies that, without any explicit time dependence $(\partial/\partial t \equiv 0)$, the energy coordinate $w$ is a constant of the motion. We note that, in accordance with the physical interpretation of the Vlasov equation, the Vlasov distribution $f({\bf x},{\bf p},t)$ satisfies the extended Vlasov equation
\begin{eqnarray}
\frac{df}{d\tau}({\bf x},{\bf p},t) &=& \{ f,\; H_{\rm e}\}_{\rm e}  \\
 &=& \frac{dt}{d\tau}\;\pd{f}{t} \;+\; \frac{d{\bf x}}{d\tau}\bdot\nabla f \;+\; \frac{d{\bf p}}{d\tau}\bdot\pd{f}{\bf p} \;\equiv\; 0, \nonumber
\end{eqnarray}
i.e., the Vlasov distribution $f({\bf x},{\bf p},t)$ is constant along particle orbits in extended phase space.

\subsection{Hamiltonian Particle Motion in Regular Phase Space} 

In their original works, Morrison \cite{Morrison_1980,Morrison_1982} \&  Marsden and Weinstein \cite{Marsden_Weinstein_1982} considered regular (noncanonical) phase space $({\bf x},{\bf p})$, for which the equations of motion \eqref{eq:xe_dot}-\eqref{eq:pe_dot} cannot be expressed in simple Hamiltonian form  \eqref{eq:Hame_eq}. However, by using the regular (noncanonical) Poisson bracket on arbitrary functions $(f,g)$ on phase space $({\bf x},{\bf p})$:
\begin{equation} 
\{ f,\; g\} \;\equiv\; \left(\nabla f\bdot\pd{g}{\bf p} \;-\; \pd{f}{\bf p}\bdot\nabla g\right) \;+\; \frac{e}{c}\,{\bf B}\bdot\left(\pd{f}{\bf p}\btimes\pd{g}{\bf p}\right),
\label{eq:PB}
\end{equation}
the equations of motion \eqref{eq:xe_dot}-\eqref{eq:pe_dot} can instead be written as
\begin{eqnarray}
\frac{dz^{\alpha}}{dt} &=& \left\{ z^{\alpha},\; K\right\} \;+\; e\,{\bf E}\bdot\left\{{\bf x},\; z^{\alpha}\right\} \nonumber \\
 &\equiv& \left\{ z^{\alpha},\; \fd{\cal H}{f}\right\} \;+\; 4\pi\,e\,\fd{\cal H}{\bf E}\bdot\left\{{\bf x},\frac{}{} z^{\alpha}\right\}.
\label{eq:z_dot}
\end{eqnarray}
Here, the contributions from the kinetic energy $K$ and the electric field ${\bf E}$ are separately expressed in terms of functional derivatives of the Vlasov-Maxwell Hamiltonian functional 
\begin{eqnarray}
{\cal H}[f,{\bf E},{\bf B}] &\equiv& \int f({\bf x},{\bf p},t)\;K({\bf p})\; d^{3}{\bf x}\,d^{3}{\bf p} \label{eq:Ham_func} \\
 &&+\; \int \left( |{\bf E}|^{2}({\bf x},t) \;+\frac{}{} |{\bf B}|^{2}({\bf x},t)\right) \frac{d^{3}{\bf x}}{8\pi},
\nonumber
\end{eqnarray}
where summation over charged particle species is implied wherever appropriate. Hence, the regular Vlasov equation for $f({\bf x},{\bf p},t)$:
\begin{eqnarray}
\pd{f}{t}({\bf x},{\bf p},t) &=& -\;\frac{d{\bf x}}{dt}\bdot\nabla f \;-\; \frac{d{\bf p}}{dt}\bdot\pd{f}{\bf p} \label{eq:f_H} \\
 &\equiv& -\;\left\{ f,\; \fd{\cal H}{f}\right\} \;-\; 4\pi\,e\,\fd{\cal H}{\bf E}\bdot\left\{{\bf x},\frac{}{} f\right\},
\nonumber
\end{eqnarray}
and the Maxwell equations for ${\bf E}({\bf x},t)$ and ${\bf B}({\bf x},t)$:
\begin{eqnarray}
\pd{\bf E}{t} &=& c\,\nabla\btimes{\bf B} \;-\; 4\pi\,e\int f\;\frac{d{\bf x}}{dt}\;d^{3}{\bf p} \label{eq:E_H}  \\
 &\equiv& 4\pi\,c\;\nabla\btimes\fd{\cal H}{\bf B} \;-\; 4\pi\,e\;\int f\;\left\{ {\bf x},\; \fd{\cal H}{f}\right\}\;d^{3}{\bf p}, 
\nonumber \\
\pd{\bf B}{t} &=& -\,c\,\nabla\btimes{\bf E} \;\equiv\; -\;4\pi\,c\;\nabla\btimes\fd{\cal H}{\bf E},
\label{eq:B_H}
\end{eqnarray}
can all be expressed in terms of  functional derivatives of the Vlasov-Maxwell Hamiltonian functional \eqref{eq:Ham_func} with respect to $(f,{\bf E},{\bf B})$: $\delta{\cal H}/\delta f = K$, $\delta{\cal H}/\delta{\bf E} = {\bf E}/4\pi$, and $\delta{\cal H}/\delta{\bf B} = {\bf B}/4\pi$.

\subsection{Vlasov-Maxwell Hamiltonian Theory} 

We now directly construct the Hamiltonian structure of the Vlasov-Maxwell equations \eqref{eq:f_H}-\eqref{eq:B_H}. For this purpose, we consider the time evolution of an arbitrary Vlasov-Maxwell functional ${\cal F}[f,{\bf E},{\bf B}]$: 
\begin{widetext}
\begin{eqnarray}
\pd{\cal F}{t} &\equiv& \int \pd{f}{t}\;\fd{\cal F}{f}\;d^{3}{\bf x}\,d^{3}{\bf p} \;+\; \int \left( \pd{\bf E}{t}\bdot\fd{\cal F}{\bf E} \;+\; \pd{\bf B}{t}\bdot\fd{\cal F}{\bf B} \right) d^{3}{\bf x}  \nonumber \\
 &=& -\;\int \fd{\cal F}{f}\; \left( \left\{ f,\; \fd{\cal H}{f}\right\} \;+\; 4\pi\,e\,\fd{\cal H}{\bf E}\bdot\left\{ {\bf x},\frac{}{} f \right\}\right) d^{3}{\bf x}\,d^{3}{\bf p} \label{eq:F_ham} \\
  &&+ \int \fd{\cal F}{\bf E}\bdot\left( 4\pi\,c\,\nabla\btimes\fd{\cal H}{\bf B} - 4\pi\,e \int f \left\{{\bf x},\; \fd{\cal H}{f}\right\}\;d^{3}{\bf p} \right) d^{3}{\bf x} - 4\pi c\,\int \fd{\cal F}{\bf B}\bdot\nabla\btimes\fd{\cal H}{\bf E}\;d^{3}{\bf x},  \nonumber
\end{eqnarray}
which can thus be expressed in Hamiltonian form in terms of  functional derivatives of the Vlasov-Maxwell Hamiltonian functional \eqref{eq:Ham_func}:
\begin{equation}
\pd{\cal F}{t} \;\equiv\; \left[ {\cal F},\frac{}{} {\cal H}\right],
\end{equation}
where the antisymmetric Vlasov-Maxwell functional bracket  \cite{Morrison_1982,Marsden_Weinstein_1982} is defined as
\begin{eqnarray}
\left[ {\cal F},\frac{}{} {\cal G}\right] &\equiv& \int f\; \left\{ \fd{\cal F}{f},\; \fd{\cal G}{f}\right\} \;d^{3}{\bf x}\,d^{3}{\bf p} + 4\pi\,e \int f \left(\fd{\cal G}{\bf E}\bdot\left\{{\bf x},\; \fd{\cal F}{f}\right\} -
\fd{\cal F}{\bf E}\bdot\left\{{\bf x},\; \fd{\cal G}{f}\right\} \right) \;d^{3}{\bf x}\,d^{3}{\bf p} \nonumber \\
 &&+\; 4\pi c \int \left( \fd{\cal F}{\bf E}\bdot\nabla\btimes\fd{\cal G}{\bf B} \;-\; \fd{\cal G}{\bf E}\bdot\nabla\btimes\fd{\cal F}{\bf B}\right) d^{3}{\bf x}.
 \label{eq:Ham_Bracket}
\end{eqnarray}
\end{widetext}
In addition, the Vlasov-Maxwell functional bracket \eqref{eq:Ham_Bracket} satisfies the Jacobi property
\[ \left[ {\cal F},\frac{}{} [{\cal G},\; {\cal H}]\,\right] \;+\; \left[ {\cal G},\frac{}{} [{\cal H},\; {\cal F}]\,\right] \;+\; \left[ {\cal H},\frac{}{} [{\cal F},\; {\cal G}]\,\right] \;\equiv 0, \]
which holds for arbitrary Vlasov-Maxwell functionals $({\cal F},{\cal G},{\cal H})$ provided the magnetic field ${\bf B}$ satisfies the divergenceless condition $\nabla\bdot{\bf B} = 0$ (see App.~B of Ref.~\cite{Morrison_2013} for a proof of the Jacobi property for the Hamiltonian bracket \eqref{eq:Ham_Bracket}). This condition is inherited from the condition for which the noncanonical Poisson bracket \eqref{eq:PB} satisfies the Jacobi property $\{f,\,\{g,\, h\}\} + \{g,\,\{h,\, f\}\} + \{h,\,\{f,\, g\}\}  = 0$.

\subsection{\label{sec:Anti_VM}Antisymmetry of the Vlasov-Maxwell Bracket}

We now note that the antisymmetry property of the Vlasov-Maxwell functional bracket \eqref{eq:Ham_Bracket} depends crucially on the expressions \eqref{eq:z_dot} and \eqref{eq:f_H}, where the contributions from the kinetic energy $K \equiv \delta{\cal H}/\delta f$ and the electric field ${\bf E} \equiv 4\pi\,\delta{\cal H}/\delta{\bf E}$ must be explicitly separated. Hence, even though the particle dynamics has a simple Hamiltonian form \eqref{eq:Hame_eq} in eight-dimensional extended phase space $({\bf x},{\bf p}; t,w)$, the Hamiltonian formulation of the Vlasov-Maxwell equations \eqref{eq:f_H}-\eqref{eq:B_H} is expressed in terms of a Hamiltonian functional \eqref{eq:Ham_func} and a Vlasov-Maxwell functional bracket \eqref{eq:Ham_Bracket}, whose antisymmetric structure requires considering particle Hamiltonian dynamics in regular phase space $({\bf x},{\bf p})$, defined in Eq.~\eqref{eq:z_dot}.

  \section{\label{sec:App}Guiding-center Phase-space Transformation}

 In this Appendix, we present the derivation of the guiding-center kinetic energy \eqref{eq:K_gc_def}, which is obtained through the guiding-center transformation in extended phase space (which includes energy $w$ and time $t$ as conjugate canonical coordinates \cite{RGL_1981}). Here, the transformation proceeds from the local particle phase-space coordinates $z^{\alpha} \equiv ({\bf x},p_{\|} \equiv {\bf p}\bdot\bhat, J_{0} \equiv m|{\bf p}_{\bot} - {\bf p}_{E}|^{2}/2\Omega,\zeta_{0};w,t)$, defined in a frame moving with the $E\times B$ momentum ${\bf p}_{\rm E} = e\,{\bf E}\btimes \bhat/\Omega$, to the guiding-center coordinates $Z^{\alpha} \equiv ({\bf X},P_{\|},J,\zeta;W,t)$, in terms of which we obtain the guiding-center Lagrangian \eqref{eq:Lag_gc}, with the guiding-center kinetic energy \eqref{eq:K_gc_def}, as they appear in Ref.~\cite{Cary_Brizard_2009}. In this drifting frame, the perpendicular particle kinetic momentum ${\bf q}_{\bot} \equiv {\bf p}_{\bot} - {\bf p}_{E}$ is assumed to be explicitly dependent on the local gyroangle $\zeta_{0}$, while $J_{0}$ is the lowest-order gyroaction that is canonically conjugate to $\zeta_{0}$.

\subsection{Guiding-center transformation in extended phase space}

We begin with the extended Poincar\'{e}-Cartan one-form in the local drifting frame \cite{RGL_1981}
\begin{eqnarray}
\gamma &=& \left[ \frac{e}{c}\,{\bf A} \;+\; \epsilon\,\left( p_{\|}\,\bhat \;+\; {\bf p}_{\rm E} \;+\frac{}{} {\bf q}_{\bot}\right)\right]\bdot\exd{\bf x} - w\,\exd t \nonumber \\
 &\equiv& \gamma_{0} \;+\; \epsilon\,\gamma_{1},
\label{eq:gamma_def}
\end{eqnarray}
while the extended phase-space Hamiltonian is
\begin{eqnarray}
H &=& e\,\Phi \;-\; w \;+\; \frac{\epsilon}{2m}\left( p_{\|}^{2} + |{\bf p}_{E}|^{2} + |{\bf q}_{\bot}|^{2} + 2\,{\bf p}_{E}\bdot{\bf q}_{\bot}\right) \nonumber \\
 &\equiv& H_{0} \;+\; \epsilon\,H_{1}.
\label{eq:Ham_def}
\end{eqnarray}
Here, the dimensionless ordering parameter $\epsilon$ is introduced through the mass renormalization $m \rightarrow \epsilon\,m$ \cite{Brizard_1995,Brizard_2023comment}, which is meant to represent the mass-over-charge ratio $m/e$ used as an expansion parameter in standard guiding-center theory \cite{Cary_Brizard_2009}.

 We now proceed with the  transformation to the extended guiding-center phase-space coordinates $Z^{\alpha} \equiv z^{\alpha} + \epsilon\,G_{1}^{\alpha} + \cdots$, where $\Gamma_{{\rm gc}0} = (e/c)\,{\bf A}\bdot\exd{\bf X} - W\,\exd t$ and $H_{{\rm gc}0} = e\,\Phi - W$ at lowest order in $\epsilon$ \cite{RGL_1981}. Next, the standard Lie-transform perturbation theory \cite{Brizard_2024} yields the first-order expressions
\begin{eqnarray}
\Gamma_{{\rm gc}1} &=& \left(P_{\|}\,\bhat \;+\; {\bf p}_{\rm E} \;+\frac{}{} {\bf q}_{\bot}\right)\bdot\exd{\bf X} \;-\; \left(\frac{e}{c}\,{\bf B}\btimes G_{1}^{\bf X}\right)\bdot\exd{\bf X} \nonumber \\
 &&+\; \left( G_{1}^{W} \;+\; \frac{e}{c}\pd{\bf A}{t}\bdot G_{1}^{\bf X}\right) \exd t, \label{eq:Gamma_1} \\
 H_{{\rm gc}1} &=& \frac{P_{\|}^{2}}{2m} \;+\; \mu\,B \;+\; \frac{|{\bf p}_{E}|^{2}}{2m} \;+\; \frac{{\bf p}_{E}}{m}\bdot{\bf q}_{\bot} \nonumber \\
  &&-\; e\,G_{1}^{\bf X}\bdot\nabla\Phi \;+\; G_{1}^{W},
 \label{eq:Ham_1}
 \end{eqnarray}
 where the first-order components $G_{1}^{\bf X}$ and $G_{1}^{W}$ are primarily chosen to eliminate the gyroangle $\zeta$-dependence introduced by ${\bf q}_{\bot}$, with the lowest-order definition $\mu \equiv |{\bf q}_{\bot}|^{2}/(2m B)$ for the guiding-center magnetic moment. In our earlier work \cite{Brizard_1995,Brizard_2024}, where we adopted the symplectic polarization representation, we selected a guiding-center transformation that kept the $E\times B$ kinetic momentum ${\bf p}_{\rm E}$ in the guiding-center Poincar\'{e}-Cartan one-form \cite{Pfirsch_1984,Pfirsch_Morrison_1985}. In the present work, however, we adopt the Hamiltonian polarization representation, and we remove ${\bf p}_{E}$ from the right side of Eq.~\eqref{eq:Gamma_1}, so that the first-order guiding-center Poincar\'{e}-Cartan one-form becomes
 \begin{equation}
 \Gamma_{{\rm gc}1} \;\equiv\; P_{\|}\,\bhat\bdot\exd{\bf X} 
 \end{equation}
 and we find the first-order components
 \begin{eqnarray}
 G_{1}^{\bf X} &=& \left({\bf p}_{\rm E} \;+\frac{}{} {\bf q}_{\bot}\right)\btimes\frac{c\bhat}{eB} \;\equiv\; -\,\left(\vb{\rho}_{\rm E} \;+\frac{}{} \vb{\rho}_{\bot}\right), \\
 G_{1}^{W} &=& -\;\frac{e}{c}\pd{\bf A}{t}\bdot G_{1}^{\bf X} \;=\; \left(\vb{\rho}_{\rm E} \;+\frac{}{} \vb{\rho}_{\bot}\right)\bdot\frac{e}{c}\,\pd{\bf A}{t}.
 \end{eqnarray}
When these components are inserted into Eq.~\eqref{eq:Ham_1}, we obtain the first-order guiding-center Hamiltonian
 \begin{eqnarray}
 H_{{\rm gc}1} &=& \frac{P_{\|}^{2}}{2m} \;+\; \mu\,B \;+\; \frac{|{\bf p}_{E}|^{2}}{2m}  \;-\; \vb{\rho}_{\rm E}\bdot e{\bf E} \nonumber \\
  &&+\; \frac{{\bf p}_{E}}{m}\bdot{\bf q}_{\bot} \;-\; \vb{\rho}_{\bot}\bdot e{\bf E} \nonumber \\
   &=&  \frac{P_{\|}^{2}}{2m} \;+\; \mu\,B \;-\; \frac{|{\bf p}_{E}|^{2}}{2m} \;\equiv\; K_{\rm gc},
   \label{eq:Ham_gc1}
 \end{eqnarray}
where we used $({\bf p}_{E}/m)\bdot{\bf q}_{\bot} =\vb{\rho}_{\bot}\bdot e{\bf E}$ and $|{\bf p}_{E}|^{2}/m =\vb{\rho}_{E}\bdot e{\bf E}$, which eliminates the gyroangle-dependence on the right side of Eq.~\eqref{eq:Ham_gc1} and introduces the negative sign in front of the last term in the guiding-center kinetic energy $K_{\rm gc}$. The additional components $(G_{1}^{P_{\|}},G_{1}^{J},G_{1}^{\zeta})$, which are obtained at higher order in $\epsilon$, will not be needed in what follows. The Reader may refer to our recent work \cite{Brizard_2024} for additional details.

\subsection{Guiding-center dynamics in extended phase space}

 As a result of this guiding-center transformation, we obtain the extended guiding-center (egc) Lagrangian
\begin{equation}
L_{\rm egc} \;=\; \frac{e}{c}{\bf A}^{*} \bdot\frac{d{\bf X}}{d\tau} - w\,\frac{dt}{d\tau} \;-\; \left(e\,\Phi + K_{\rm gc} \;-\frac{}{} w\right),
\label{eq:Lag_egc}
\end{equation}
where $\tau$ denotes the orbit parameter in extended phase space, the guiding-center kinetic energy $K_{\rm gc}$ is given by Eq.~\eqref{eq:K_gc_def}, and we use the definition $(e/c)\,{\bf A}^{*} = (e/c)\,{\bf A} + P_{\|}\,\bhat$. From this extended Lagrangian, we obtain the 
Euler-Lagrange equations in extended guiding-center phase space $Z^{a} = ({\bf X},P_{\|};w,t)$:
\begin{eqnarray}
\frac{dP_{\|}}{d\tau}\,\bhat &=& e\,{\bf E}^{*} \;-\; \nabla K_{\rm gc} \;+\; \frac{e}{c}\frac{d{\bf X}}{d\tau}\btimes{\bf B}^{*}, \label{eq:Xgc_EL} \\
\bhat\bdot\frac{d{\bf X}}{d\tau} &=& \pd{K_{\rm gc}}{P_{\|}} \;=\; P_{\|}/m, \label{eq:Pgc_EL} \\
\frac{dt}{d\tau} &=& 1, \label{eq:wgc_EL} \\
\frac{dw}{d\tau} &=& \pd{}{t}\left(e\,\Phi \;+\frac{}{} K_{\rm gc}\right) \;-\; \frac{e}{c}\,\pd{{\bf A}^{*}}{t}\bdot\frac{d{\bf X}}{d\tau}, \label{eq:tgc_EL}
\end{eqnarray}
where ${\bf E}^{*} \equiv {\bf E} - e^{-1}P_{\|}\,\partial\bhat/\partial t$ and ${\bf B}^{*} \equiv \nabla\btimes{\bf A}^{*}$.

Next, from the extended Euler-Lagrange equations \eqref{eq:Xgc_EL}-\eqref{eq:Pgc_EL}, we obtain the guiding-center equations of motion
\begin{eqnarray}
\frac{d{\bf X}}{d\tau} &=& \frac{P_{\|}}{m}\,\frac{{\bf B}^{*}}{B_{\|}^{*}} + \left(e\,{\bf E}^{*} \;-\frac{}{} \nabla K_{\rm gc}\right)\btimes\frac{c\bhat}{eB_{\|}^{*}}, \label{eq:X_gc_e} \\
\frac{dP_{\|}}{d\tau} &=& \left(e\,{\bf E}^{*} \;-\frac{}{} \nabla K_{\rm gc}\right)\bdot\frac{{\bf B}^{*}}{B_{\|}^{*}}, \label{eq:P_gc_e}
\end{eqnarray}
where $B_{\|^{*}} \equiv \bhat\bdot{\bf B}^{*}$. These equations can also be expressed in Hamiltonian form 
\begin{equation}
dZ^{a}/d\tau \;=\; \{Z^{a}, H_{\rm egc}\}_{\rm egc}
\label{eq:Ham_egc}
\end{equation}
in terms of the extended guiding-center Hamiltonian $H_{\rm egc} \equiv e\,\Phi + K_{\rm gc} - w$ and the extended Poisson bracket
\begin{eqnarray}
\{ F,\; G\}_{\rm egc} &\equiv& \pd{F}{w}\,\pd{G}{t} - \pd{F}{t}\,\pd{G}{w} \nonumber \\
 &&+\; \frac{{\bf B}^{*}}{B_{\|}^{*}}\bdot\left(\nabla^{*}F\;\pd{G}{P_{\|}} \;-\; \pd{F}{P_{\|}}\;\nabla^{*}G \right) \nonumber \\
 &&-\; \frac{c\bhat}{eB_{\|}^{*}}\bdot\nabla^{*}F\btimes\nabla^{*}G,
 \label{eq:egc_PB}
 \end{eqnarray}
 where
 \[ \nabla^{*} \;\equiv\; \nabla \;-\; \frac{e}{c}\,\pd{{\bf A}^{*}}{t}\;\pd{}{w}, \]
 so that $e\,{\bf E}^{*} - \nabla K_{\rm gc} = -\,\nabla^{*}H_{\rm egc}$. The reduced guiding-center equations of motion \eqref{eq:X_gc_e}-\eqref{eq:P_gc_e} can, of course, be expressed in terms of the reduced guiding-center Poisson bracket  \eqref{eq:gcPB}, as shown in Eqs.~\eqref{eq:Xgc_dot}-\eqref{eq:Pgc_dot}. As discussed in Sec.~\ref{sec:Anti_VM}, the separation of $K_{\rm gc} \equiv \delta{\cal H}_{\rm gc}/\delta F_{\rm gc}$ and ${\bf E}^{*} \equiv 4\pi\,\delta^{\star}{\cal H}_{\rm gc}/\delta{\bf D}_{\rm gc}$ plays a crucial role in guaranteeing the antisymmetry of the guiding-center Hamiltonian functional bracket \eqref{eq:gcVM_bracket}.

\bibliographystyle{unsrt}
\bibliography{gc_extended-gcVM}

\end{document}